\documentclass[final]{appolb}
\usepackage{graphicx}

\begin{document}

\title{
  NEUTRINO MASSES FROM {\sl R}-PARITY NON-CONSERVING LOOPS
  \thanks{Talk presented at the Cracow Epiphany Conference on Neutrinos
    and Dark Matter, 5--8.1.2006, Cracow, Poland}}

\author{
  Marek G\'o\'zd\'z, Wies{\l}aw A. Kami\'nski
  \address{
    Department of Informatics, Maria Curie-Sk{\l}odowska University, \\
    pl. Marii Curie--Sk{\l}odowskiej 5, 20-031 Lublin, Poland}
  \and
  Fedor \v Simkovic
  \address{
    Institute f\"ur Theoretische Physik der Universit\"at
    T\"ubingen, \\ D-72076 T\"ubingen, Germany \\
    Department of Nuclear Physics, Comenius University, \\
    Mlynsk\'a Dolina F1, SK--842 15 Bratislava, Slovakia}
}

\maketitle

\begin{abstract}
  We present new formulae for the neutrino masses generated by
  $R$-parity violating interactions within minimal supersymmetric
  standard model. The importance of inclusion of $CP$ phases in the
  neutrino mass matrix is discussed in detail.
\end{abstract}


\section{Introduction}

The existence of physics beyond the Standard Model (SM) has been
suspected for a~long time, but only quite recently this hypothesis
gained a~strong experimental evidence. The observation of oscillations
of the solar, atmospheric, and reactor types of neutrinos
\cite{SK,SNO,Kamland,K2K} confirmed the non-zero mass of these
particles, thus shedding light on new physics.

The study of neutrino properties, both experimental and theoretical, is
of primary importance for the development of non-standard physics. Most
effects predicted by various models beyond the SM are extremely weak and
therefore very difficult to observe in experiments. The possibility of
collecting precise neutrino data is the best chance by now to get an
insight into this range of physics.

In the present paper we investigate the model of Majorana neutrino masses
generated by particle--sparticle loops within supersymmetric standard
model without $R$-parity \cite{loops,mg-loops,mg-new}. In particularly
we discuss the influence of $CP$ Dirac and Majorana phases on the
phenomenological neutrino mass matrix elements used in the calculations.

\section{Neutrino masses from trilinear $R$-parity violating
  interactions}

Let us briefly recall the basics of the minimal supersymmetric standard
model (MSSM) \cite{mssm}. The model is described by the superpotential
$W+W^{\not R}$, where the so-called $R$-parity conserving part of the
superpotential has the form
\begin{equation}
  W = \epsilon_{ab} [(\mathbf{Y}_E)_{ij} L_i^a H_1^b \bar E_j
  + (\mathbf{Y}_D)_{ij} Q_i^{ax} H_1^b \bar D_{jx}
  + (\mathbf{Y}_U)_{ij} Q_i^{ax} H_2^b \bar U_{jx} + \mu H_1^a H_2^b ],
\end{equation}
while its $R$-parity violating part reads
\begin{equation}
  W^{\not R} = \epsilon_{ab}
  \left[
    \frac{1}{2} \lambda_{ijk} L_i^a L_j^b \bar E_k
    + \lambda'_{ijk} L_i^a Q_j^{xb} \bar D_{kx} + \kappa^i L_i^a H_2^b 
  \right]
  + \frac{1}{2}\epsilon_{xyz} \lambda''_{ijk}\bar U_i^x\bar
  D_j^y \bar D_k^z. 
\end{equation}
Our notation is as follows: {\bf Y}'s are 3$\times$3 Yukawa matrices,
$L$ and $Q$ stand for lepton and quark left-handed $SU(2)$ doublet
superfields while $\bar E$, $\bar U$ and $\bar D$ denote the
right-handed lepton, up-quark and down-quark $SU(2)$ singlet
superfields, respectively. $H_1$ and $H_2$ mean two Higgs doublet
superfields. We have introduced color indices $x,y,z = 1,2,3$,
generation indices $i,j,k=1,2,3$ and the SU(2) spinor indices $a,b,c =
1,2$.

The $R$-parity is an accidental symmetry present in the MSSM, and is
defined as $R=(-1)^{3B+L+2S}$, where $B$, $L$, and $S$ are the baryon,
lepton, and spin numbers, respectively. The introduction of $R$-parity
violation $W^{RpV}$ implies the existence of lepton or baryon number
violating processes, like the unobserved proton decay and neutrinoless
double beta decay ($0\nu2\beta$). In order to get rid of too rapid
proton decay and to allow for lepton number violating processes it is
customary to set $\lambda''=0$.

\begin{figure}[h]
  \centering
    \includegraphics[width=0.45\textwidth]{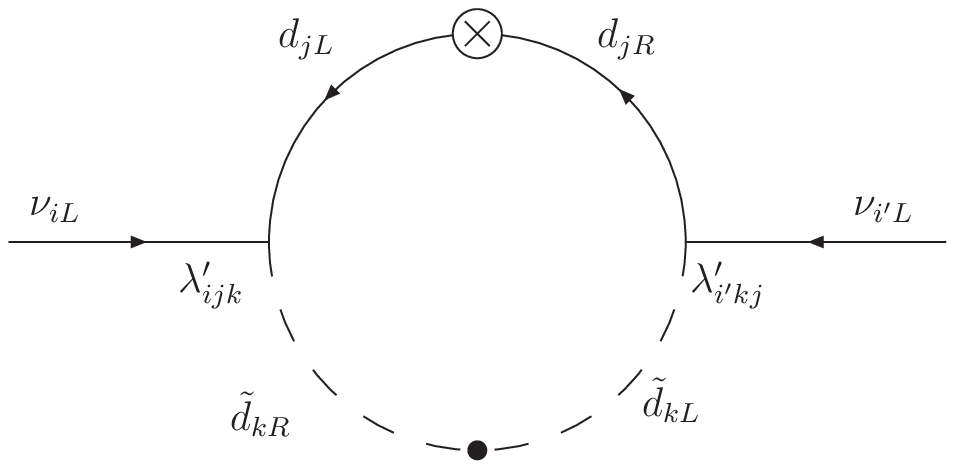}
    \includegraphics[width=0.45\textwidth]{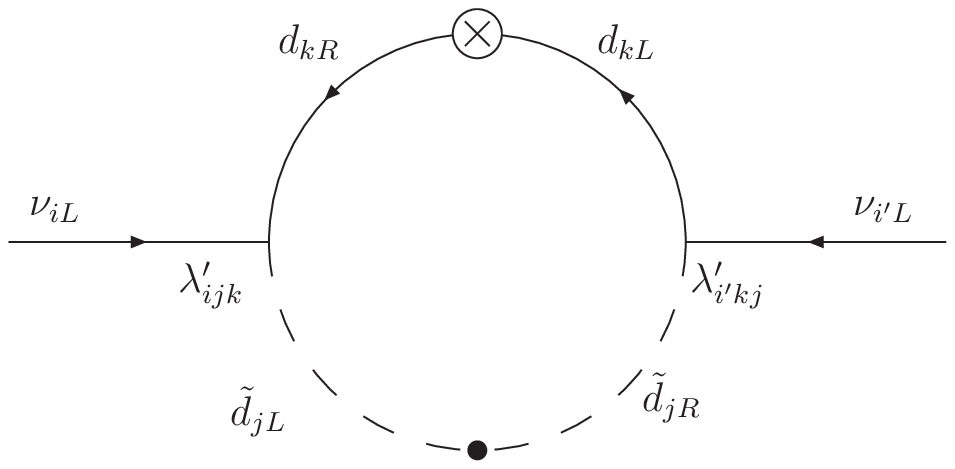} \\
    \includegraphics[width=0.45\textwidth]{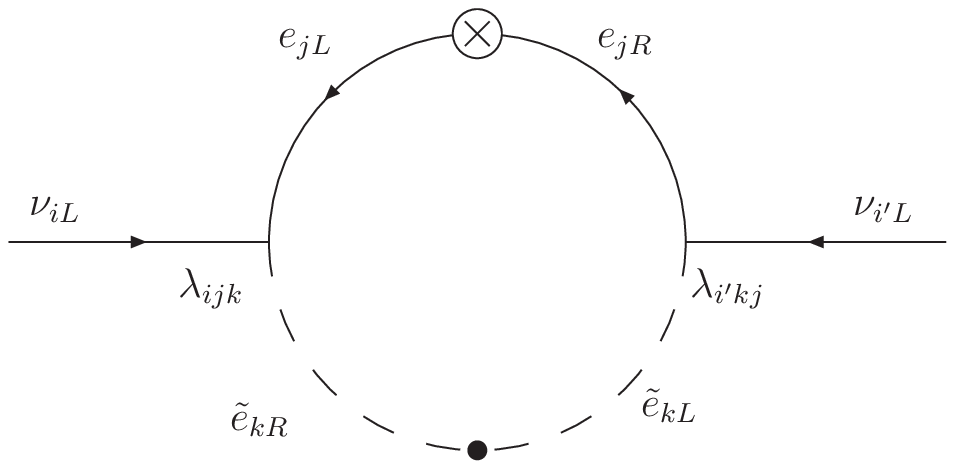}
    \includegraphics[width=0.45\textwidth]{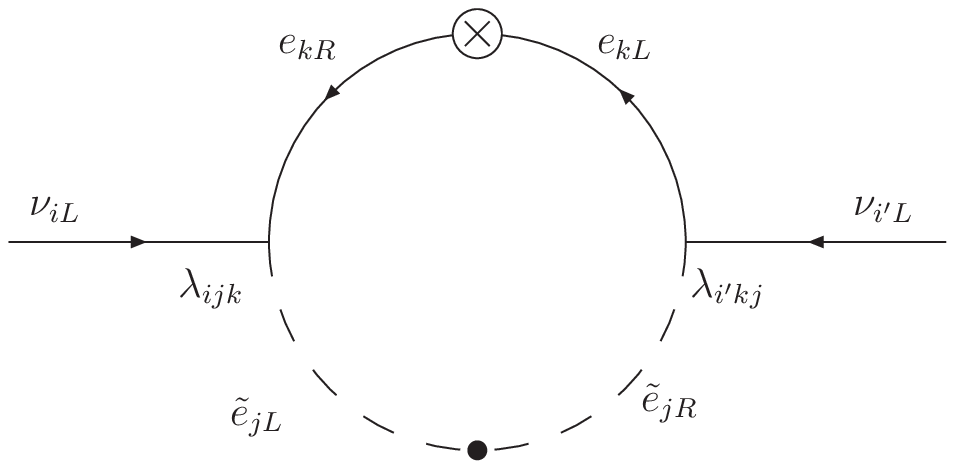}
    \caption{\label{fig1} Feynman diagrams leading to Majorana neutrino
      masses.}
\end{figure}

The $R$-parity violating interactions imply the existence of Feynman
diagrams depicted on Fig.~\ref{fig1}, which generate the one-loop
Majorana neutrino mass terms. The amplitudes of these processes give the
following shape of the mass matrices. From the quark--squark
contribution we have:
\begin{equation} 
  {\cal M}_{ii'}^q =
  \frac{3}{16\pi^2} \sum_{jkla} \left \{ 
    \lambda'_{ijk}\lambda'_{i'kl} \left (
      V_{ja} V_{la} v^q_{ak} m_{d^a} \right ) 
  + \lambda'_{ijk}\lambda'_{i'lj} \left (
      V_{ka} V_{la} v^q_{aj} m_{d^a} \right ) \right \},
\label{Mqq}
\end{equation}
where the loop integral is
\begin{eqnarray}
  v^q_{jk} = \frac{\sin(2\theta^k)}{2} \left(
    \frac{\log x_2^{jk}}{1-x_2^{jk}} - 
    \frac{\log x_1^{jk}}{1-x_1^{jk}} \right ).
\label{vq}
\end{eqnarray}
Here $m_{d^j}$ is $j$-th generation down quark mass, $\theta^k$ is the
squark mixing angle between the $k$-th squark mass eigenstates
$M_{\tilde d^k_{1,2}}$, and $x_{1,2}^{jk} = m_{d^j}^2 / M_{\tilde
  d^k_{1,2}}^2$. The factor 3 comes from summation over three colors of
quarks. Eq.~(\ref{Mqq}) includes the possible quark mixing through the
CKM matrix $V$.

For loops containing lepton--slepton pairs we do not include the weak
effect of mixing of leptons. The resulting neutrino mass matrix reads:
\begin{equation}
  {\cal M}_{ii'}^\ell = 
  \frac{1}{16\pi^2} \sum_{jk} \lambda_{ijk}\lambda_{i'kj} 
  (v^\ell_{jk} m_{e^j} + v^\ell_{kj} m_{e^k}),
\label{Mll}
\end{equation}
with the loop integral $v^\ell$ having analogous form to $v^q$ with
quarks and squarks replaced by leptons and sleptons, respectively.

\section{Phenomenological mass matrices and $CP$ phases}

The presented above formulae Eqs. (\ref{Mqq})--(\ref{Mll}) contain
superparticle masses and trilinear coupling constants $\lambda$ and
$\lambda'$. The masses of superparticles in MSSM may be generated using
the renormalization group equations, with some unification conditions
assumed at high energies (see \cite{mg-loops} and references therein).
It is important to note, that we have no information about the $\lambda$
and $\lambda'$ couplings neither at $m_{Planck}$ nor at $m_Z$ and
therefore we cannot use the RGE procedure for them (although the
required RG equations are known). It follows that, after obtaining the
numerical forms of ${\cal M}^q$ and ${\cal M}^\ell$ from the
experimental data, we can put constraints on the non-standard couplings
$\lambda$ and $\lambda'$. This idea has already been discussed in
literature in various forms \cite{loops}. All of them, however, were
based on many simplifications, like the approximate treatment of squark
(slepton) mixing, negligence of quark mixing and very simplified
treatment of the MSSM mass spectrum.

A~problem directly related to the discussed mass formulae is the
determination of the phenomenological neutrino mass matrices from the
experimental data. These can be evaluated using the well-known relation
${\cal M}^{ph} = U\cdot diag(m_1,m_2,m_3)\cdot U^T$, $m_i$ being the
neutrino mass eigenvalues. The standard parameterization of the PMNS
matrix in terms of the three mixing angles is:
\begin{equation}
  U = \left (
  \begin{array}{ccc}
    c_{12} c_{13} & s_{12} c_{13} & s_{13} e^{-i \delta} \\
    -s_{12} c_{23} - c_{12} s_{23} s_{13} & c_{12} c_{23} - s_{12}
    s_{23} s_{13} & s_{23} c_{13} \\
    s_{12} s_{23} - c_{12} c_{23} s_{13} & -c_{12} s_{23} - s_{12}
    c_{23} s_{13} & c_{23} c_{13}
  \end{array}
  \right ),
  \label{U}
\end{equation}
where $s_{ij} = \sin\theta_{ij}$, $c_{ij} = \cos\theta_{ij}$, and
$\theta_{ij}$ is the mixing angle between the neutrino flavor
eigenstates labeled by indices $i$ and $j$. The recent global analysis
of neutrino oscillations \cite{maltoni} yields the best fit values:
$\sin\theta_{12}~=~0.55$, $\sin\theta_{23}~=~0.71$ and
$\sin\theta_{13}~=~0$.

\begin{figure}[h]
  \centering
  \vskip 2 true mm
    \includegraphics[width=0.85\textwidth]{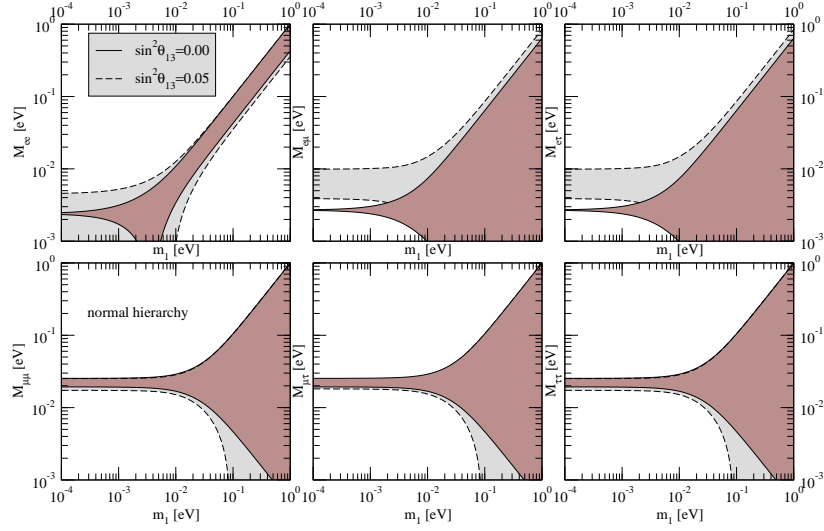}
    \caption{\label{fig2} Allowed values of the neutrino mass matrix
      elements as functions of $m_1$ in the case of normal neutrino mass
      hierarchy.  The best fit values of mixing angles were used with
      the exception of $\sin^2\theta_{13}$ for which we consider two
      separate cases.}
\end{figure}

\begin{figure}[h]
  \centering
  \vskip 4 true mm
    \includegraphics[width=0.85\textwidth]{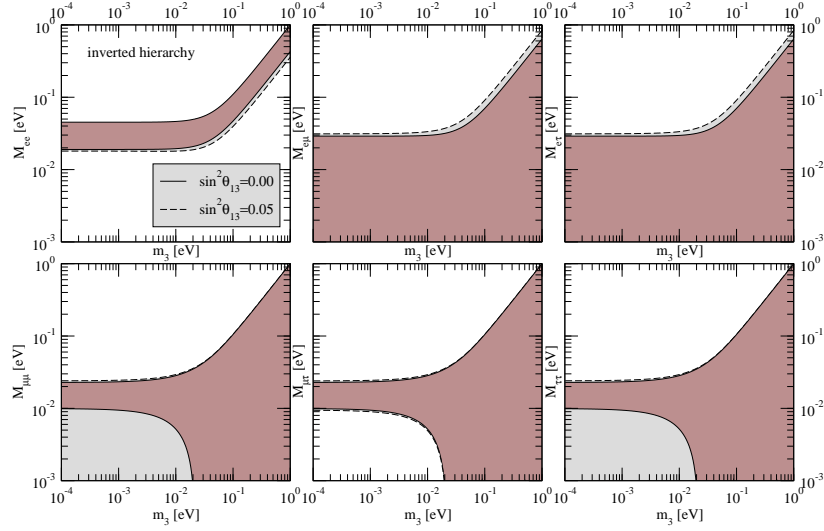}
    \caption{\label{fig3} Same as Fig.~\ref{fig2} but assuming the
      inverted hierarchy of neutrino masses.}
\end{figure}

In the case of Majorana neutrinos the matrix (\ref{U}) has to be
multiplied by $diag(1,e^{i \alpha_{21}},e^{i \alpha_{31}})$. The
Majorana phases $\alpha_{21}$ and $\alpha_{31}$, and the Dirac phase
$\delta$ are undetermined. They can take arbitrary values and for
simplicity are often neglected. This simplification may lead to
significant errors. Let us investigate the influence of all three phases
on the elements of ${\cal M}^{ph}$ in the case of normal hierarchy of
neutrino masses. In such a~case we have $m_{1} \ll m_{2} \ll m_{3}$, and
all the mass eigenvalues may be expressed in terms of the lightest
$m_1$: $m_{2} = \sqrt{\Delta m^{2}_{{21}} + m_1^2}$ and $m_{3} =
\sqrt{\Delta m^{2}_{{31}} + m_1^2}$, where $\Delta m^{2}_{{21}}=7.1
\times 10^{-5}$ eV$^2$, and $\Delta m^{2}_{{31}}=2.0 \times 10^{-3}$
eV$^2$ \cite{maltoni}. The results are presented on Fig.~\ref{fig2}.
A~similar analysis for the case of inverted hierarchy is shown on
Fig.~\ref{fig3}.

One sees that the influence of the possible $CP$ phases is much greater
than the uncertainty in the experimental values of the neutrino mixing
angles. This proves the importance of careful treatment of these phases
in theoretical calculations. At this point one remark is in order. We
have found several regions of $m_{1(3)}$ for which there exist
combinations of phases which give certain elements in the neutrino mass
matrix equal to zero.  This is an interesting observation, especially
when related to the ${\cal M}_{ee}$ element, which is governing the
$0\nu2\beta$ decay rate.  It turns out that for some sets of parameters
this decay may be suppressed even for three massive Majorana neutrinos.
For example, our calculations suggest that ${\cal M}_{ee} = 0$ for
normal hierarchy and $5.72\cdot 10^{-5} < m_1 < 7.94\cdot 10^{-3}$ eV.

As a~consequence for the trilinear $R$-parity violating mechanism
described in the previous section, these combinations of neutrino
parameters imply either $\lambda=\lambda'=0$, or the existence of other
mechanism which suppresses the amplitudes of the diagrams from
Fig.~\ref{fig1}. A~more detailed study on this topic is surely needed.

\section{Conclusions}

We have discussed the possibility of generating small Majorana neutrino
masses within the framework of $R$-parity violating minimal
supersymmetric standard model. The 1-loop level contributions are given
by particle--sparticle loops. We have calculated the amplitudes of such
processes taking into account quark, squark and slepton mixing in an
exact manner. We have also outlined the procedure for finding
constraints on the trilinear non-standard coupling constants $\lambda$
and $\lambda'$. The numerical results will be published in a~full-length
paper \cite{mg-new}. We have also discussed in detail the importance of
$CP$ phases present in the phenomenological neutrino mass matrix, which
can have an important impact on the $R$-parity violating MSSM model as
well as on the predictions for exotic nuclear processes.

{\bf Acknowledgments.} MG greatly acknowledges the financial support
from the Foundation for Polish Science. FS acknowledges the support of
the EU ILIAS project (RII3-CT-2004-506222) and the VEGA Grant agency of
the Slovak Republic (No.~1/0249/03).


\end{document}